# A Novel Transfer Learning Approach for Mental Stability Classification from Voice Signal

**Rafiul Islam**
4IR Research Cell
Daffodil International University, Dhaka, Bangladesh
rafiul15-14353@diu.edu.bd

**Dr. Md. Taimur Ahad**
Associate Professor and Associate Head
Department of CSE
Daffodil International University, Dhaka, Bangladesh
Taimurahad.cse@diu.edu.bd

# A Novel Transfer Learning Approach for Mental Stability Classification from Voice Signal

## Abstract

This study presents a novel transfer learning approach and data augmentation technique for mental stability classification using human voice signals and addresses the challenges associated with limited data availability. Convolutional neural networks (CNNs) have been employed to analyse spectrogram images generated from voice recordings. Three CNN architectures, VGG16, InceptionV3, and DenseNet121, were evaluated across three experimental phases: training on non-augmented data, augmented data, and transfer learning. This proposed transfer learning approach involves pre-training models on the augmented dataset and fine-tuning them on the non-augmented dataset while ensuring strict data separation to prevent data leakage. The results demonstrate significant improvements in classification performance compared to the baseline approach. Among three CNN architectures, DenseNet121 achieved the highest accuracy of 94% and an AUC score of 99% using the proposed transfer learning approach. This finding highlights the effectiveness of combining data augmentation and transfer learning to enhance CNN-based classification of mental stability using voice spectrograms, offering a promising non-invasive tool for mental health diagnostics.

# 1. Introduction

Mental health disorders are a major public health problem, with millions of people affected around the world. According to the World Health Organization (WHO), one in four people will have a mental health issue at some point in their life, which in turn highlights the need for effective assessments and interventions (Mental Health, 2022). Conventional evaluation methods, such as self-reported questionnaires and clinical interviews, are subjective and prone to underdiagnosis or misdiagnosis, particularly in the presence of mental health-related stigma (Prizeman et al., 2023). However, these methods are time-consuming and require trained professionals, limiting accessibility to these methods, especially in low-resource settings (Jordans et al., 2019). This has led to an increasing interest in exploring the possibility of using technology to enhance mental health assessment, including deep learning to analyse voice signals (Biswas et al., 2024).

Voice signal analysis is a promising method for assessing mental health. Research indicates that vocal characteristics, such as pitch, tone, and speech patterns, can provide crucial insights into mental health (Bhat et al., 2023). Changes in these vocal parameters are linked to conditions such as depression and anxiety, suggesting that voice analysis could be a reliable mental health indicator (Cummins et al., 2018).

The objectives of this study are threefold: first, to evaluate the performance of various deep convolutional neural network (CNN) models on non-augmented datasets; second, to assess the impact of data augmentation techniques on model performance; and third, to apply a novel transfer learning approach that utilises pre-trained models on augmented datasets before testing on non-augmented datasets to address dataset limitations and improve classification efficiency.

Recent advancements in deep learning, particularly the application of CNNs to spectrogram images, have shown promise in enhancing classification accuracy across various domains, including audio event classification and emotion recognition (Seo et al., 2022; Qi et al., 2017). CNNs are particularly effective because they can automatically extract hierarchical features from images, which is crucial when dealing with complex data, such as voice spectrograms (Khamparia et al., 2019).

Data augmentation techniques are essential for improving the robustness of the model, particularly when working with limited datasets (Slizovskaia et al., 2022). By artificially increasing the diversity of training data through transformations such as pitch shifting, time stretching, and noise addition, researchers can enhance the generalisability of their models

(Amer, 2023). This is particularly relevant in voice signal analysis, where variations in recording conditions and speaker characteristics can significantly impact model performance (Wubet & Lian, 2022).

Transfer learning, an advanced deep learning technique, allows models to use knowledge from one domain to improve performance in another related domain (Iman et al., 2023; Essa, 2023). This is especially useful when labelled data are limited, such as in mental health datasets. This method addresses data scarcity and leverages knowledge from related tasks to improve target task performance. However, concerns regarding data leakage during transfer learning highlight the need for transparency and methodological rigour (Lam, 2024).

This study aims to address the following key questions:

> RQ1) How does data augmentation affect mental stability detection using voice signal spectrogram images?
>
> RQ2) Which type of deep CNN performs better in detecting and classifying mental stability using voice signal spectrogram images?
>
> RQ3) Does the proposed transfer-learning approach improve the performance of the deep CNN models?
>
> RQ4) Can transfer learning be effectively implemented when data availability is limited to pre-training the model?

This study aims to contribute to the field of mental health assessment by answering these questions, identifying the best-performing CNN architecture for this task, and showing the benefits of data augmentation and transfer learning techniques in improving the model's performance.

The key contributions of this study are as follows.

i. This study employed a custom-collected dataset from mental health institutions in Bangladesh, offering valuable data that enhances the understanding of mental health diagnostics in a culturally specific context.

ii. This research highlights strict data separation and subject-independent partitioning to prevent data leakage and ensure the experimental results' integrity and validity.

iii. This study evaluated the impact of data augmentation techniques on model performance by analysing both augmented and non-augmented datasets, thereby providing practical guidance on how data augmentation can influence the robustness of mental health classification models.

iv. This research conducted a thorough comparative analysis of various deep CNN architectures to identify the most effective model for classifying mental stability from voice spectrogram images.

v. This study introduces a new approach to transfer learning, in which a model is pre-trained on an augmented version of a voice spectrogram dataset and then fine-tuned on the raw version of the same dataset.

vi. This research addresses the uses of a transfer learning approach with limited data availability.

## 2. Literature review

**2.1 Deep CNNs for Mental Health Detection**

Recent research has shown that deep learning models, particularly Convolutional Neural Networks (CNNs), have been increasingly used to analyse voice signals, particularly for tasks related to mental health diagnostics. The application of CNNs in voice-based mental stability studies has demonstrated notable success rates, with models achieving high accuracy in classifying emotional states and detecting mental health disorders using audio signals (Bagheri & Power, 2022; Shin et al., 2017). Several deep learning architectures can be applied to spectrogram analysis, each with its strengths in different aspects of feature extraction. Khan et al. (2019) conducted an extensive survey of deep CNN architectures and classified them based on their architectural innovations. Key CNN architectures include the following.

**Spatial Exploitation-Based CNNs**

Models such as VGG16 focus on spatial hierarchy extraction, which is crucial for capturing time–frequency relationships in spectrograms. These models have been widely used in image recognition tasks and have shown promise in voice analysis owing to their ability to capture fine-grained details in spectrograms. For instance, VGG16 has been successfully used to classify voice quality deviations, demonstrating its effectiveness in distinguishing between normal and pathological voice signals (Uloza et al., 2015).

**Depth-Based CNNs**

InceptionV3 introduces deeper architectures that use techniques such as residual connections and multiscale processing. These models are particularly effective in extracting complex patterns from spectrograms, making them ideal for analysing subtle variations in voice that may indicate mental health conditions. Research has shown that InceptionV3's multi-path architecture allows it to capture a wide range of features simultaneously, which is crucial for voice analysis, where variations in pitch and tone can be critical indicators of mental stability (Zhou et al., 2019).

**Multi-Path-Based CNNs**

DenseNet121 focuses on feature reuse, which can be beneficial for learning from nuanced variations in voice spectrograms. Their efficient use of parameters and ability to capture intricate voice features make them suitable for mental stability classification. Studies have indicated that DenseNet121 outperforms traditional CNN architectures in various applications, including medical image classification and voice pathology detection, owing to its ability to learn rich feature representations (Farahani, 2017).

These architectures provide the foundation for modern spectrogram-based analysis in mental health detection, and their comparative performance has yet to be explored in the context of detecting mental stability from voice.

## 2.2 Data Augmentation Impact on Spectrogram Analysis

Data augmentation has proven to be a critical technique for improving the generalisation of deep-learning models, particularly when working with limited datasets. In voice signal analysis, augmentation methods, such as pitch shifting, time stretching, and noise injection, enhance model performance by artificially increasing the diversity of training datasets (Salamon & Bello, 2017). For instance, studies have demonstrated that augmenting audio data can improve the robustness of classification tasks, allowing models to generalise better to unseen data (Chlap et al., 2021; Handoyo et al., 2022). The effectiveness of data augmentation is particularly relevant in mental health diagnostics, where obtaining large, labelled datasets can be challenging owing to privacy concerns and the sensitive nature of the data (Park & Caragea, 2020). By employing data augmentation techniques, researchers can create more diverse training sets that better represent the variability found in real-world voice signals. This is crucial for developing models that accurately classify mental stability across different populations and contexts (Zhao et al., 2019).

However, the impact of specific augmentation techniques can vary significantly depending on the task's nature and the dataset's characteristics. Some augmentation methods, such as noise injection, may inadvertently degrade the model performance if not applied judiciously (Ren et al., 2021). This underscores the importance of carefully selecting and evaluating augmentation strategies to ensure that they contribute positively to model training. Moreover, although data augmentation has been extensively studied in image classification tasks, its application in voice signal analysis remains less explored, indicating a potential area for further research (Chlap et al., 2021).

## 2.3 Transfer Learning in Limited Dataset Contexts

Transfer learning has gained traction as a powerful approach for improving model performance in various domains, including voice signal analysis. Researchers can fine-tune these models on smaller, task-specific datasets by leveraging pre-trained models on large, diverse datasets, thereby enhancing the classification performance (Singh & Singh, 2020). This technique is particularly beneficial in mental health diagnostics, where obtaining large, labelled datasets can be challenging owing to privacy concerns and the sensitive nature of the data. Several studies have explored the use of transfer learning in mental health diagnostics. Toto et al. (2021) introduced AudiBERT, a deep transfer-learning framework for depression classification using voice signals. This model, which integrates a self-attention mechanism, achieved a high performance, demonstrating the potential of transfer learning in mental health applications. Similarly, Naderi et al. (2019) applied multimodal deep learning to predict mental disorders from audio speech samples, leveraging pre-trained models to combine features from both audio and text, thereby enhancing the predictive accuracy of the models. However, the effectiveness of transfer learning is often contingent on the similarity between source and target domains, which can limit its applicability in certain contexts (Zhao et al., 2019).

Asutkar and Tallur et al. (2023) proposed an innovative approach to transfer learning in the context of limited data in their study on machine fault diagnosis. They introduced a deep transfer learning strategy in which models were pre-trained on public datasets with lab-generated datasets to enhance domain generalisation (Asutkar & Tallur, 2023). Their methodology addresses the challenge of limited labelled data by leveraging data augmentation to enrich the training set. This enables the model to learn robust features that generalise new data well. This approach is particularly relevant in our context, where data scarcity constrains the availability of large and diverse datasets.

Inspired by their success, this study adopted a similar strategy but within the domain of mental health classification using voice spectrograms. By pre-training on an augmented version of the training data and fine-tuning the raw data, we aim to maximise the utility of the available data and improve the model's performance despite dataset limitations. This approach also emphasises the prevention of data leakage, which is a critical concern in transfer learning. Data leakage can occur when models are trained on test data or when there is an overlap between training and testing datasets, leading to overly optimistic performance estimates (Lam, 2024). The proposed methodology maintains the integrity of the evaluation process by ensuring a strict separation between the augmented training data and validation/test datasets.

## 3. Methodology

The experiments in this study were conducted based on Google Collab, using the Keras library. TensorFlow, one of the best Python deep-learning libraries available for machine learning, was used in this study. DL models were trained using a Google Collab Tesla graphics processing unit (GPU). TPU is available through Google's collaborative framework. Initially, the collab framework provided up to 12 GB of random-access memory (RAM) and approximately 360 GB of GPU in the cloud for research purposes.

### 3.1 Dataset description

The primary challenge of this study was the unavailability of publicly accessible voice data from individuals with mental instability. To address this, data were collected from the Pabna Mental Hospital and the National Institute of Mental Health (NIMH) in Bangladesh, ensuring ethical compliance and confidentiality. Participants were categorised based on the clinical assessments of mentally unstable and mentally stable individuals. Voice recordings were captured using high-quality equipment in controlled environments and preprocessed by reducing background noise, normalising volume, and segmenting audio into two-second clips. These clips were converted into spectrogram images using Short-Time Fourier Transform (STFT) with specific parameters (48,000 Hz sampling rate, 128 mel bands, 2048 window size,

512 hop length, and 8000 kHz maximum frequency). The mathematical representation of this Short-Time Fourier Transform (STFT) can be expressed as

$$X(t,f) = \sum_{n=-\infty}^{\infty} x[n] \cdot w[n-t] \cdot e^{-j2\pi fn} \tag{1}$$

where $x[n]$ represents the audio signal, $w[n-t]$ denotes the window function and $t$ and $f$ denote the time and frequency, respectively.

Figure 1 shows sample spectrogram images generated from the audio recordings. These images visualise the frequency components of the audio signals and offer insights into the spectral characteristics of the audio tapes. The X-axis represents different time frames measured in seconds, while the y-axis indicates the amplitude, showing the dynamic range of the recordings (see Figure 1).

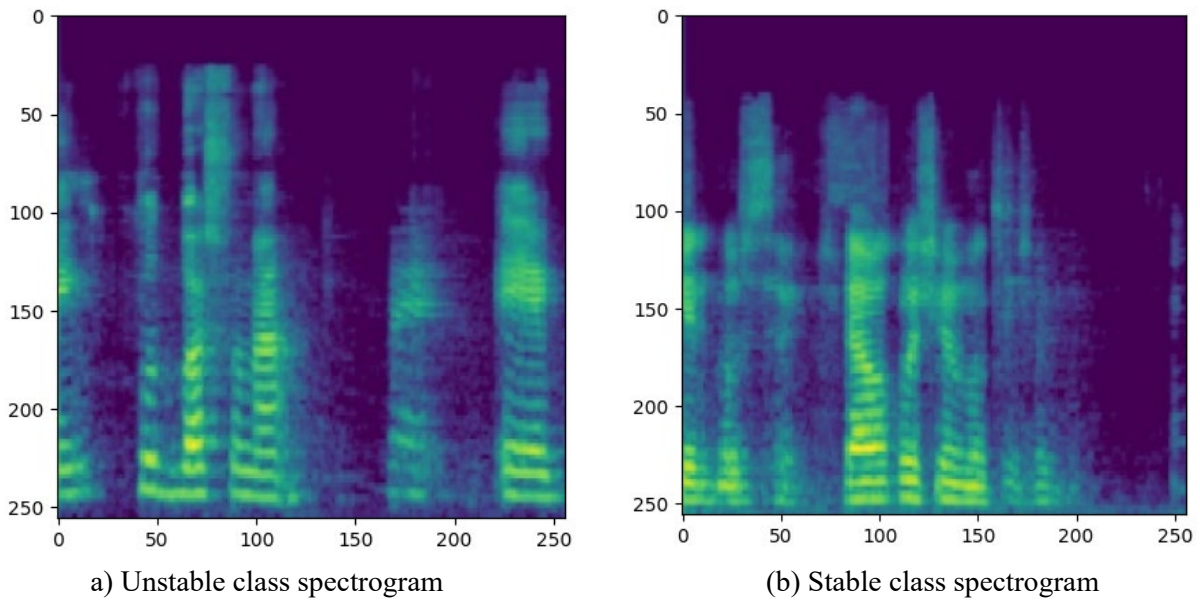

a) Unstable class spectrogram          (b) Stable class spectrogram

Figure 1: Spectrogram images from Unstable and Stable Classes

These spectrogram images balance time and frequency resolutions and reveal specific vocal features crucial for mental stability analysis, such as pitch, tone, and frequency patterns. This study used these spectrograms to classify mental stability using deep CNN models. This dataset is publicly available in the Mendeley data repository and can be accessed through the following link: https://data.mendeley.com/datasets/s5j25b5tjk/1

## 3.2 Data Augmentation Process
### 3.2.1 Data Split

To prevent data leakage, the dataset was divided into training (70%), validation (15%), and test (15%) sets using subject-independent partitioning, to ensure that there was no overlap of samples from the same individual across sets.

### 3.2.2 Data Augmentation Pipeline

Data augmentation was applied exclusively to the training set using the Imaging Augmentation Library (IAA). The pipeline combines SpecAugment, Gaussian noise, and random erasing in random sequences with varying probabilities to enhance training data variability. Specific techniques based on Park et al. (2019) include Time Masking, which randomly removes portions of the time steps to introduce variability; Frequency Masking, which masks frequency bins to simulate different acoustic conditions; SpecAugment, which combines both time and frequency masking to enhance generalisation; Gaussian Noise, which adds random noise to simulate real-world conditions; and Random Erasing, which erases rectangular regions in the spectrogram to encourage learning from incomplete data. It is crucial to exclude the validation and test sets from this augmentation pipeline to maintain the dataset integrity and prevent performance improvements from being attributed to data leakage. Some samples of the augmented images are shown in Figure 2.

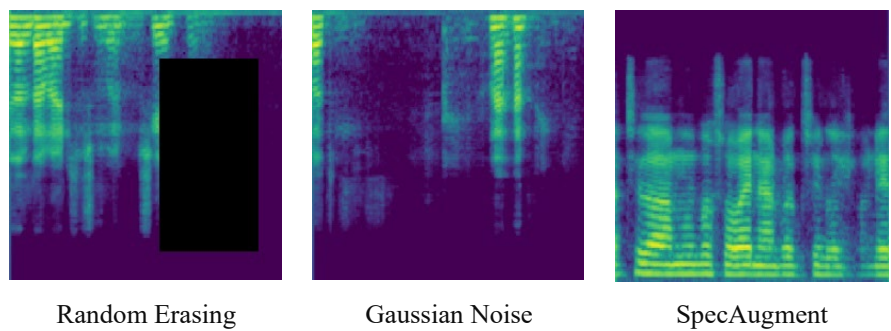

Random Erasing      Gaussian Noise      SpecAugment

Figure 2: Sample augmented images used in this study

### 3.3 Deep Learning Selection

This study chose three Convolutional Neural Network (CNN) architecture categories: Spatial Exploration-Based CNNs, Depth-Based CNNs, and Multi-Path-Based CNNs. This categorisation is predicated on architectural innovations that enhance CNNs' representational capacity of CNNs (Khan et al., 2019). Each category was designed to capture different feature extraction and processing aspects, which are crucial when analysing voice spectrograms for mental stability classification (see Figure 3).

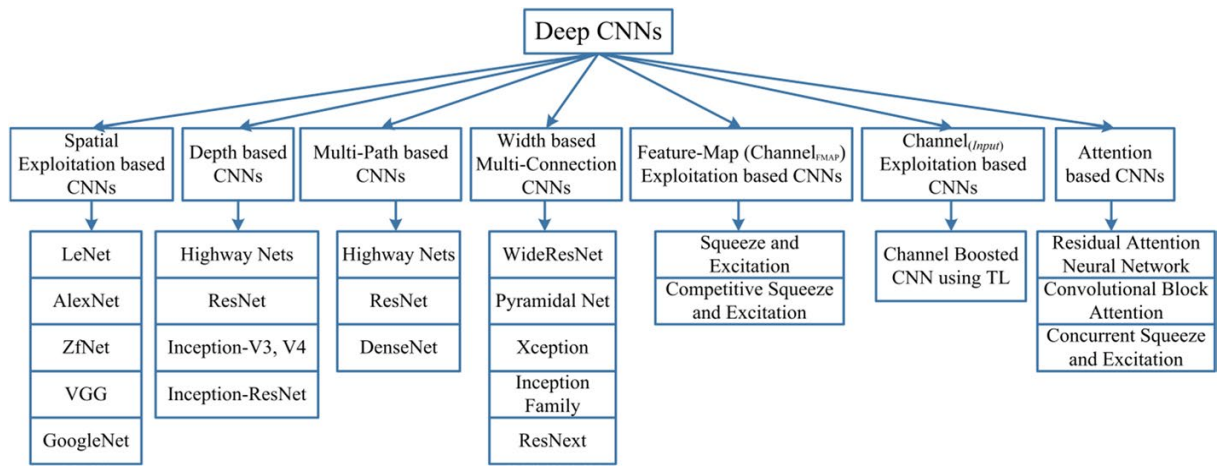

Figure 3: Deep CNN architectures showing seven different categories (Khan et al., 2019)

**Spatial Exploitation-Based CNNs (VGG16)**

Spatial exploitation-based CNNs were designed to learn spatial hierarchies within data. This is particularly useful in spectrogram analysis, where spatial relationships in the time-frequency domain are crucial. In voice spectrogram analysis, spatial patterns correspond to temporal changes in speech and frequency that are vital for distinguishing between mentally stable and unstable voice signals. VGG16 employs small receptive fields ($3 \times 3$ convolutions) and multiple layers, which contribute to the extraction of high-level features from spectrogram images. These models are particularly well-suited for this task because of their ability to learn fine-grained details and hierarchical representations, which are essential when analysing subtle variations in voice data related to mental stability.

**Depth-Based CNNs (InceptionV3)**

Depth-based CNNs increase network depth to improve feature extraction without encountering the vanishing gradient problem. InceptionV3 employs a multiscale approach with parallel convolutions to capture fine and coarse patterns, making it well-suited for analysing spectrograms with complex frequency-time relationships. This model is crucial for mental stability detection because it allows the network to capture complex speech patterns that may indicate mental instability, such as subtle variations in tone or pacing.

**Multi-Path-Based CNNs (DenseNet121)**

Multi-path-based CNNs, such as DenseNet121, aim to enhance learning efficiency and reduce redundancy by reusing features across different layers. DenseNet121 promotes maximum feature reuse and results in more robust feature extraction with fewer parameters, making it effective for spectrogram analyses. Multi-path networks, such as DenseNet121, are particularly well suited for this study because they can efficiently learn from the complex and subtle variations in voice spectrograms, which are essential for detecting mental health-related anomalies.

## 3.4 Experimental Design

**Training Configurations**

The selected CNNs were trained using a batch size of 32 for the augmented dataset and 16 for the non-augmented dataset, striking a balance between computational efficiency and manageable dataset representation per iteration. Early Stopping callbacks were used for 250 epochs, and patience was used for 10 iterations. Patience is the number of epochs with no improvement, after which the training is stopped. The Adam optimiser was employed due to its adaptive learning rate set at $\alpha = 0.0001$, $\beta1 = 0.9$, $\beta2 = 0.999$ and $\epsilon = 1 \times 10 - 7$. Simultaneously, categorical cross-entropy was utilised as the loss function to quantify the classification error for this classification task (distinguishing between stable and unstable mental health states). The same optimiser and cross-entropy were used for all the CNN models, and the models were saved .h5 files.

The experiment followed a three-phase approach to evaluate the effects of data augmentation and transfer learning on CNN performance.

**Phase 1:** In Phase 1, the chosen CNN architectures were trained and evaluated on a non-augmented dataset to establish a baseline for classification accuracy. Each model was evaluated using standard metrics, such as classification accuracy, confusion matrix, and area under the ROC curve (AUC) (see Figure 4).

**Phase 2:** Apply the same CNN architecture to the augmented dataset and compare its performance with the non-augmented baseline. After completing the training, all the model files were saved separately as .h5 files without interacting with the test data. The saved model files contain only the learned parameters from the training on the augmented data. These saved models were used in Phase 3 as pre-trained models. After saving the model files, each model was evaluated using standard metrics, such as classification accuracy, confusion matrix, and area under the ROC curve (AUC). This phase seeks to quantify the effects of data augmentation on classification accuracy and generalisation (see Figure 4).

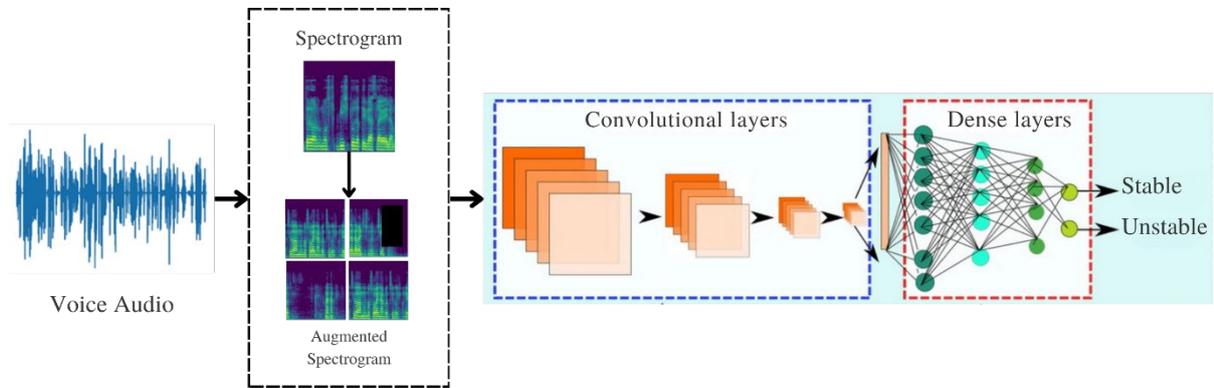

Figure 4: Workflow diagram of CNN Architectures using non-augment or augmented dataset

**Phase 3:** In Phase 3, the pre-trained models from Phase 2, which were trained on the augmented dataset, were utilised as the starting point for transfer learning. This phase involves fine-tuning the pre-trained models on a non-augmented dataset to adapt the learned features to the specific characteristics of the raw data. The primary objective of this phase is to leverage the robust feature representations acquired during pre-training on augmented data and refine them to improve the classification accuracy and generalisation of non-augmented data ( Figure 5).

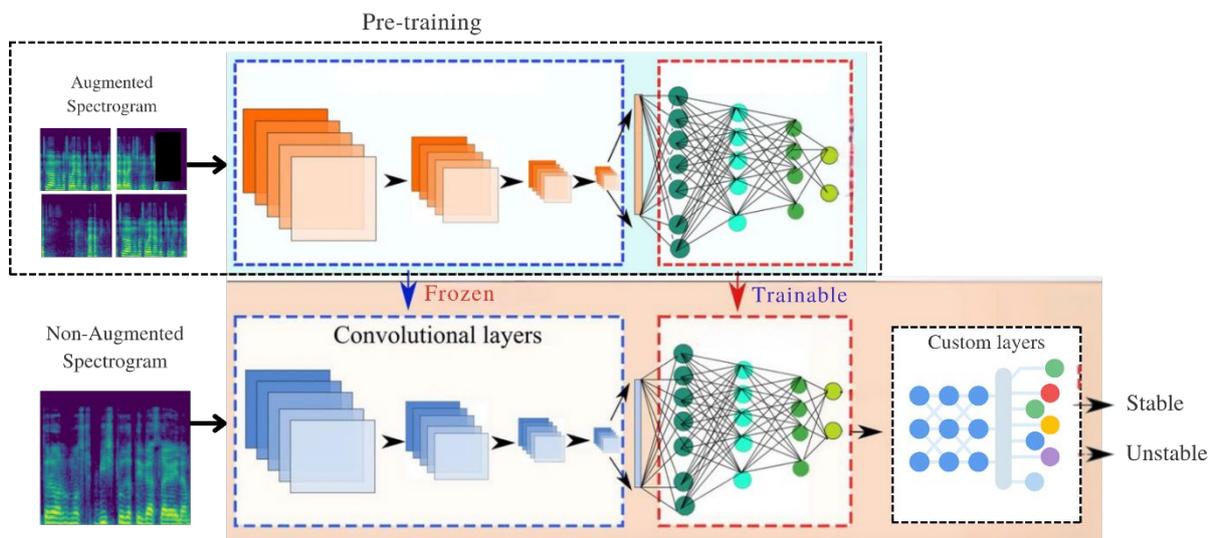

Figure 5: Workflow diagram of proposed Transfer Learning Architecture

### 3.5 Transfer learning process

**Training Strategy and Justifications**

This study proposes a novel transfer learning methodology for mental stability detection using voice spectrograms, addressing the challenges of limited data and domain relevance. Instead of traditional transfer learning, which pre-trains models on unrelated large datasets, this approach leverages the same dataset in two forms, augmented and non-augmented, to optimise model performance. Augmented data introduce transformations such as noise addition or time shifting, helping the model learn robust features from varied inputs without collecting new data. The model was pre-trained on this augmented dataset to learn generalised feature representations that capture domain variability. The raw, non-augmented dataset was then fine-tuned to adapt these features to the specific characteristics of the unaltered voice spectrograms. Data leakage was prevented by ensuring that the validation and test sets remained untouched and that the evaluation integrity was maintained. Furthermore, subject independence across dataset splits was guaranteed by including all samples from a particular subject in one split, avoiding individual-specific feature learning, which could artificially inflate performance metrics. This methodology enhances model generalisation, reduces overfitting, improves classification accuracy, and effectively transfers knowledge to non-augmented data while preserving domain consistency and preventing data leakage.

**Transfer Learning Training Procedure**

The following algorithm outlines the sequential training procedure proposed in this study.

---
**Algorithm 1: Proposed Transfer Learning Workflow**

---
**Input:** non-augmented training dataset ***Dtrain.***

**Model Initialization**

- Select and initialise the pre-trained CNN model from the previous phase.

- Randomly initialise the weights $\theta$ of the model.

**Model Loading and Layer Freezing**

- Load the pre-trained model and weights $\theta$ from pre-training using '***load_model***'.

- Set the trainable attribute of pre-trained layers to **False.**

**Add Custom Layers for Fine-tuning**

- **The** intermediate (sixth) layer output **was selected** as **input**.

- Applies a global average over spatial dimensions.

---

- Normalizes the inputs to improve training stability.

- Randomly drops a fraction *P* of units to prevent overfitting.

- Adds fully connected layers with activation functions (ReLU).

- Incorporate weight regularisation (*L2* regularisation).

- Final Dense layer with *C* (number of classes) and softmax activation:

$$softmax\ (z_i) = \frac{\exp(z_i)}{\sum_{j=1}^{C} \exp(z_j)} \quad (2)$$

- Set the trainable attribute of new layers to **True.**

**Compile the Model**

- Choose a small learning rate ($\alpha = 1 \times 10^{-5}$) For fine-tuning.

- Use an appropriate loss function *L*, such as categorical cross-entropy, for classification tasks.

- Employ optimisation algorithms (Adam).

- Specify evaluation metrics (accuracy).

**Set Up Callbacks**

- Save the model weights $\theta^*$ when the validation loss improves.

- Reduce $\alpha$ when the validation loss plateaus.

- Stop training when validation accuracy stops improving.

**Train the Model:**

- Train for 250 epochs, using callbacks to monitor performance.

**Apply Regularization Techniques**

- Use dropout layers in the new fully connected layers.

- Apply *L2* regularisation to weights in the dense layers.

**Evaluation**

- Use test set **Dtest** to evaluate model performance.

- Vary the classification threshold to compute the true positive rate (TPR) and the false positive rate (FPR).

**Output:** Classification accuracy

# 4. Results of the experiments

The experimental results are presented in this section. The experiments were conducted in three phases: training on non-augmented data to establish a baseline, training on the augmented dataset to assess the impact of augmentation, and fine-tuning of pre-trained models on non-augmented data using the transfer learning methodology. The evaluation was performed using the classification accuracy, confusion matrix, and ROC curve metrics. The results of each phase are summarised in the following sections, including figures and tables, to illustrate the performance of different CNN architectures.

## 4.1 Baseline Performance on Non-Augmented Data (Phase 1)

The initial phase of the experiment involved training various CNN architectures on a non-augmented dataset. This phase served as a baseline for comparing the effects of data augmentation and transfer learning. The tested models included VGG16, InceptionV3, and DenseNet121, which were chosen for their ability to extract hierarchical and deep features from spectrogram images. Table 1 lists the classification accuracy, precision, recall, and F1 score of each model.

Table 1: Baseline Performance on Non-Augmented Dataset

| Model | Accuracy | Precision | Recall | F1-Score |
|---|---|---|---|---|
| **VGG16** | 0.89 | 0.96 | 0.84 | 0.90 |
| **InceptionV3** | 0.88 | 0.94 | 0.84 | 0.89 |
| **DenseNet121** | 0.92 | 0.96 | 0.89 | 0.92 |

The results demonstrated that DenseNet121 achieved the highest classification accuracy of 0.92, followed by VGG16. Although VGG16 performed exceptionally well in terms of precision, its overall accuracy was slightly lower than that of DenseNet121.

**Loss-Accuracy Curve from Non-Augmented Data**

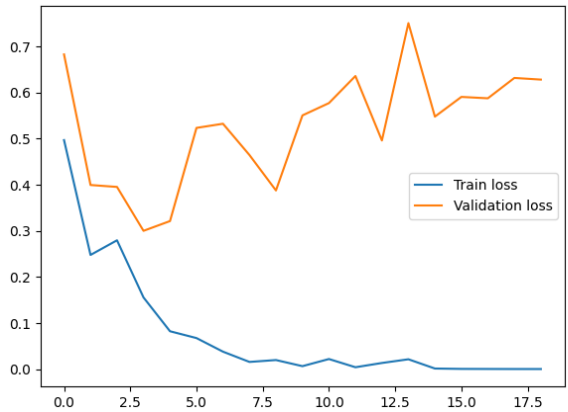
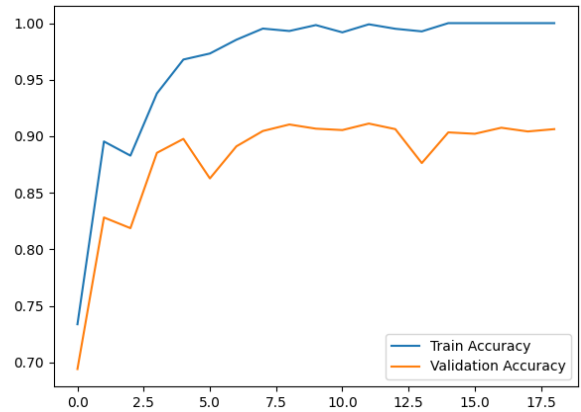

(a) VGG16

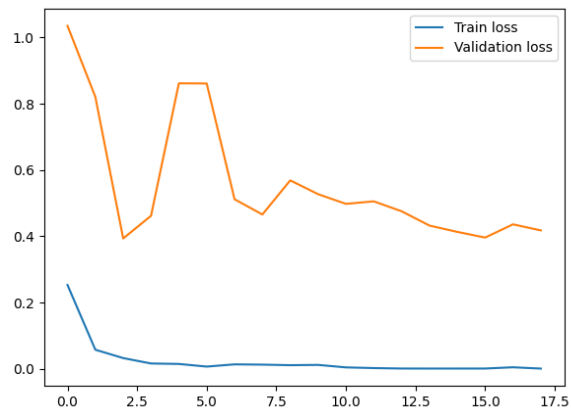
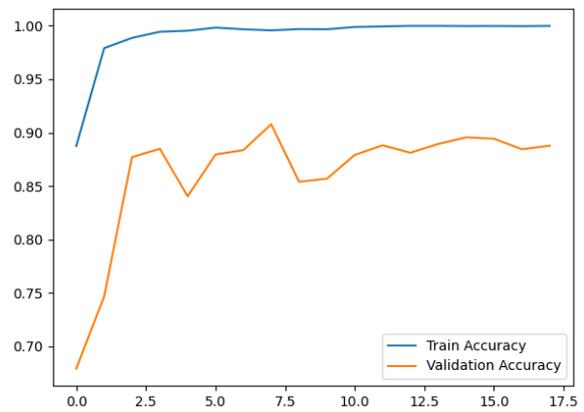

(b) InceptionV3

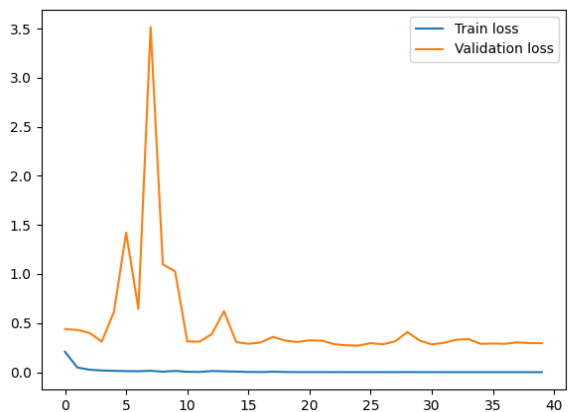
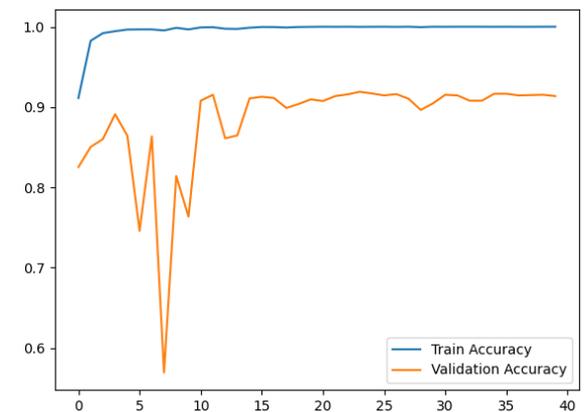

(c) DenseNet121

*Figure 6: Training-Validation Loss and Accuracy Curves for Non-Augmented Data*

Figure 6 shows that all models exhibit good convergence during training. The training and validation loss curves decreased steadily, indicating effective learning. The accuracy curves show that DenseNet121 achieved the highest validation accuracy, confirming its superior performance on the non-augmented dataset.

**Confusion Matrices from Non-Augmented Data**

The confusion matrices (Figure 7) further highlight each model's performance, showing the distribution of correct and incorrect predictions across mental stability categories.

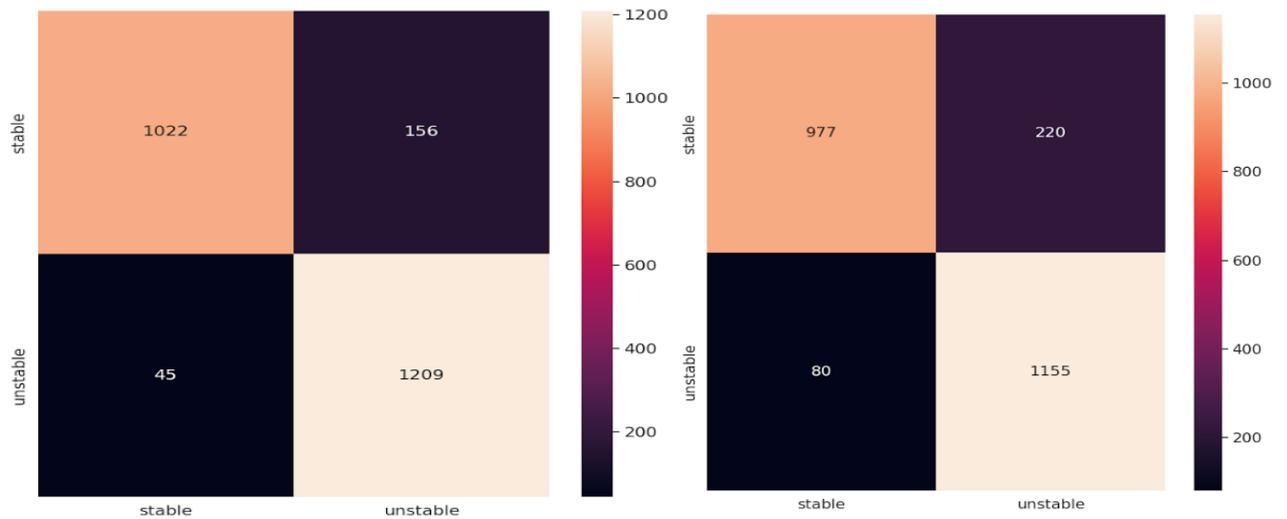

(a) VGG16            (b) InceptionV3

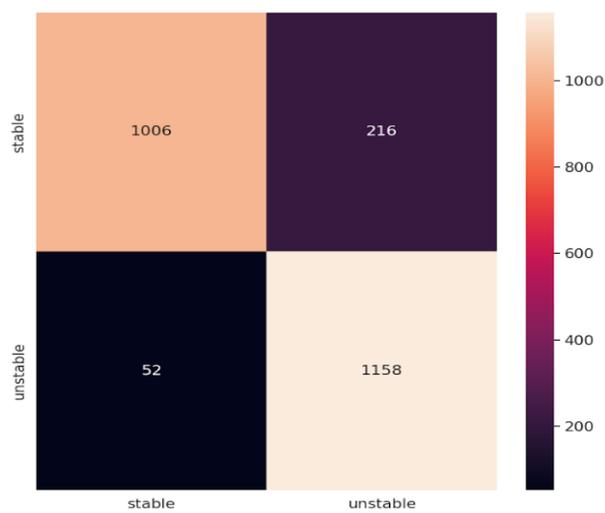

(c) DenseNet121

*Figure 7: Confusion Matrices for CNN Models on Non-Augmented Dataset*

The confusion matrices in Figure 7 illustrate the classification performance of each model. DenseNet121 correctly classified more samples into stable and unstable categories, resulting in fewer misclassifications than VGG16 and InceptionV3.

**ROC Curves from Non-Augmented Data**

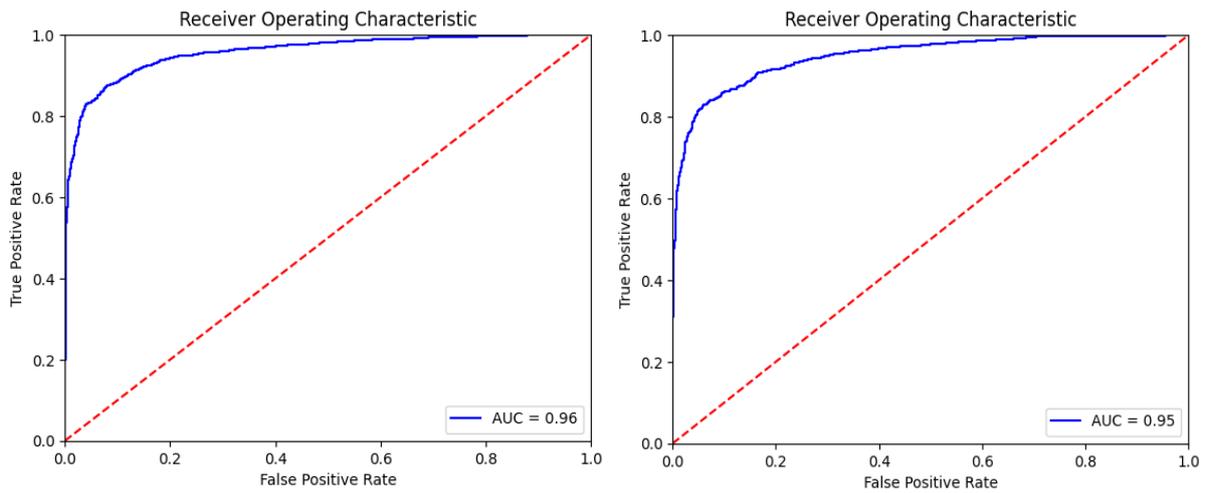

(a) VGG16  (b) InceptionV3

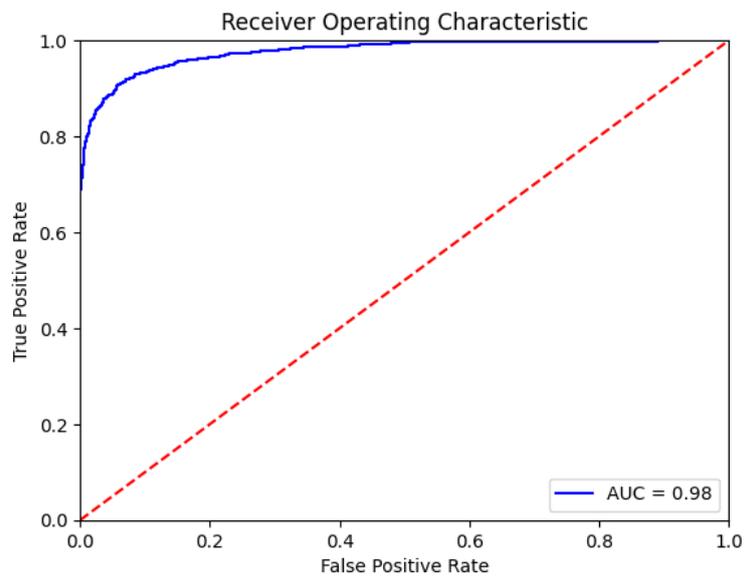

(c) DenseNet121

Figure 8: ROC Curves for CNN Models on Non-Augmented Dataset

The ROC curves in Figure 8 illustrate the trade-off between true-positive and false-positive rates for each model. DenseNet121's ROC curve was closest to the top-left corner, indicating a better discrimination ability. The AUC values are listed in Table 2.

Table 2: AUC Values for Models Trained on Non-Augmented Dataset

| Model | AUC |
| --- | --- |
| **VGG16** | 0.96 |
| **InceptionV3** | 0.95 |
| **DenseNet121** | 0.98 |

Table 2 shows that DenseNet121 achieved the highest AUC of 0.98, confirming its superior performance in classifying mental stability on the non-augmented dataset.

## 4.2 Performance on Augmented Dataset (Phase 2)

In the second phase, the same CNN architectures were trained on the augmented dataset, which included various augmentation techniques, such as speckle, Gaussian noise, and random erasing. This phase aimed to assess whether data augmentation could improve model performance by introducing variability into the training data. Table 3 lists the performance metrics of each model for the augmented dataset.

Table 3: Performance on Augmented Dataset

| Model | Accuracy | Precision | Recall | F1-Score |
| --- | --- | --- | --- | --- |
| **VGG16** | 0.90 | 0.93 | 0.90 | 0.91 |
| **InceptionV3** | 0.91 | 0.94 | 0.90 | 0.92 |
| **DenseNet121** | 0.93 | 0.95 | 0.92 | 0.94 |

After training on the augmented dataset, significant improvements were observed in all models. DenseNet121 again performed the best, achieving an accuracy of 0.93. The results indicate that data augmentation effectively improved the generalisation capabilities of the models, as evidenced by the increased accuracy, precision, and recall of each CNN model.

**Loss Accuracy Curve from Augmented Data**

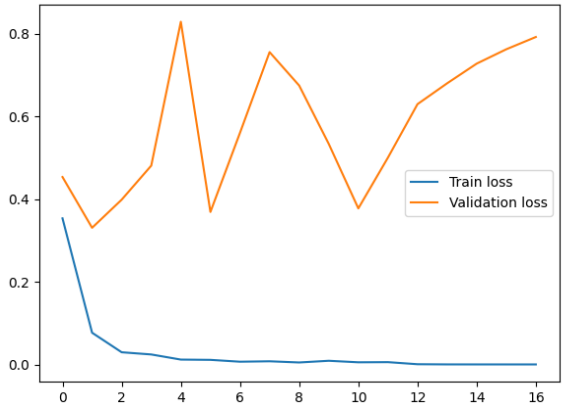 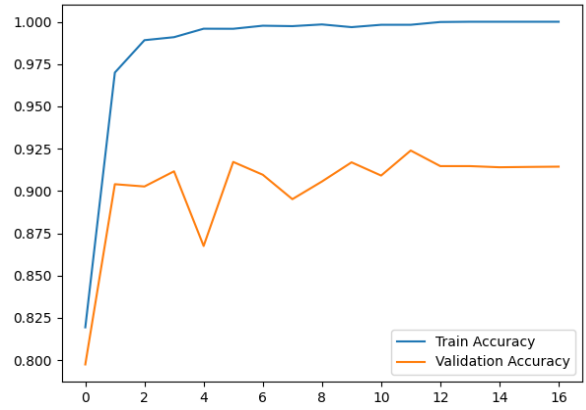

(a) VGG16

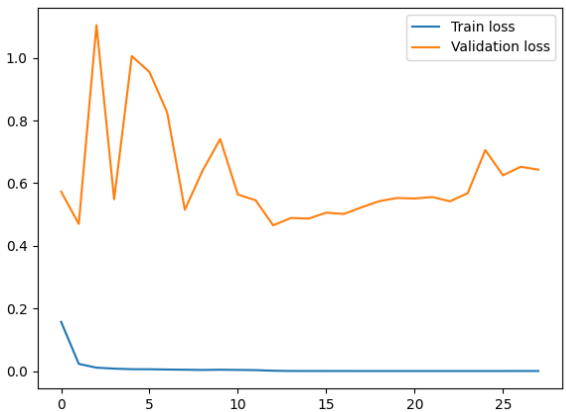 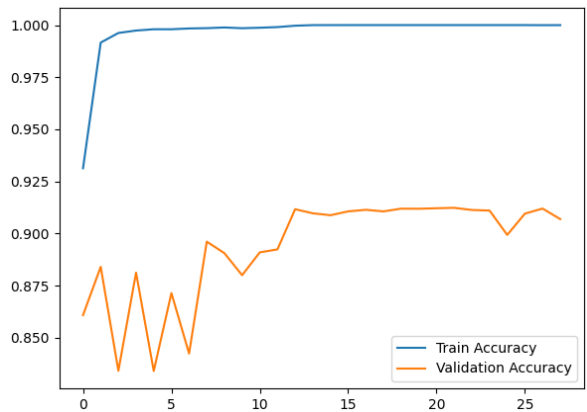

(b) InceptionV3

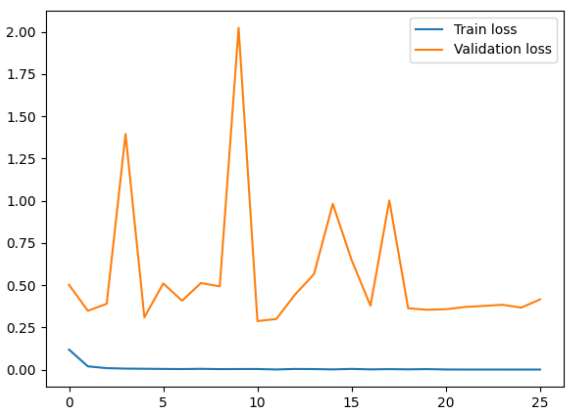 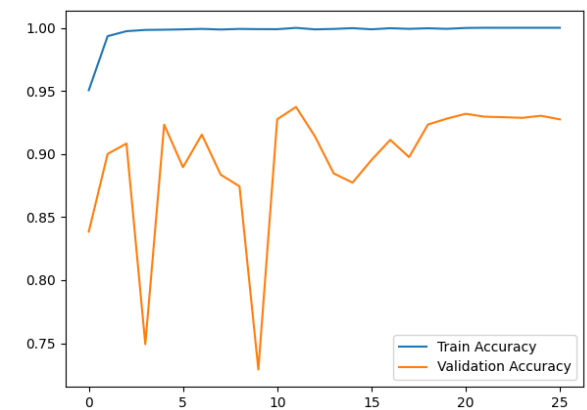

(c) DenseNet121

*Figure 9: Training-Validation Loss and Accuracy Curves for Augmented Data*

Figure 9 shows that the models achieved lower training and validation losses than those in Phase 1. The accuracy curves show improved validation accuracy, indicating that data augmentation helps prevent overfitting and enhances generalisation.

**Confusion Matrices from Augmented Data**

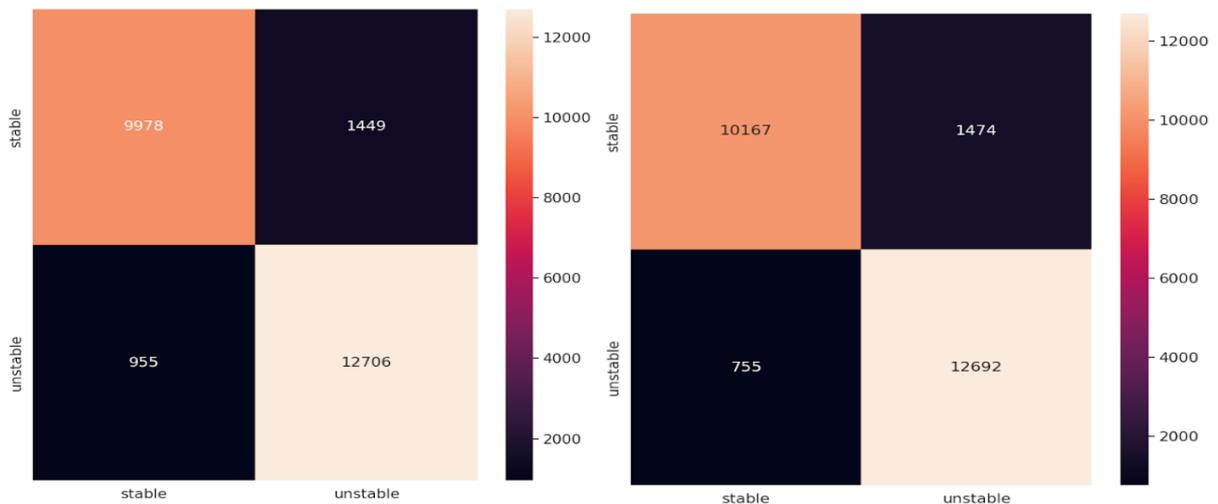

(a) VGG16  (b) InceptionV3

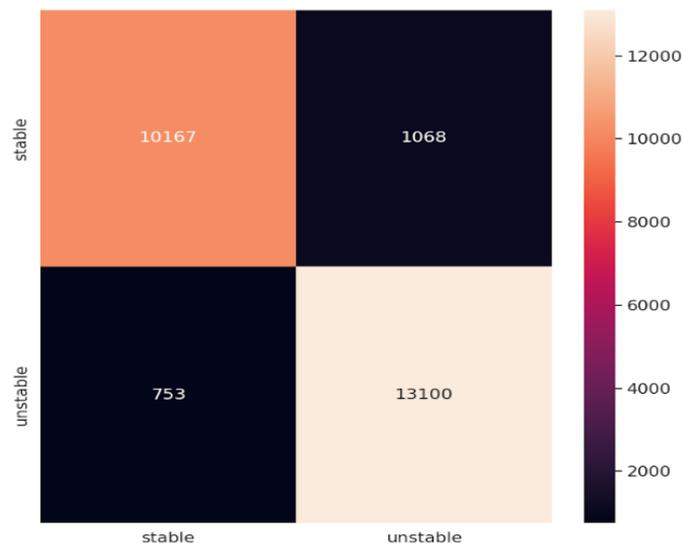

(c) DenseNet121

*Figure 10: Confusion Matrices for CNN Models on Augmented Dataset*

The confusion matrices in Figure 10 show that the number of correctly classified samples increased for all models. DenseNet121 displayed the highest number of true positives in both classes, thereby reducing misclassification errors.

**ROC Curves from Augmented Data**

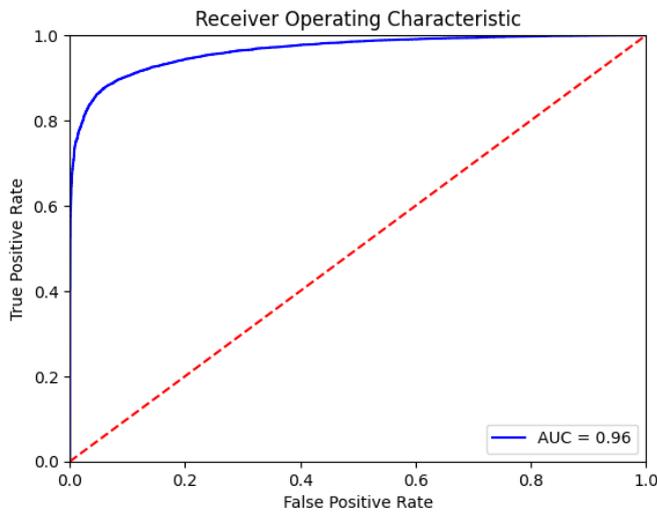

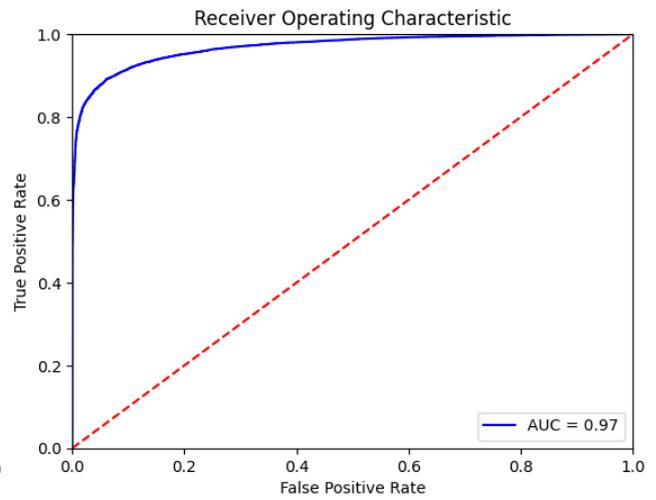

(a) VGG16   (b) InceptionV3

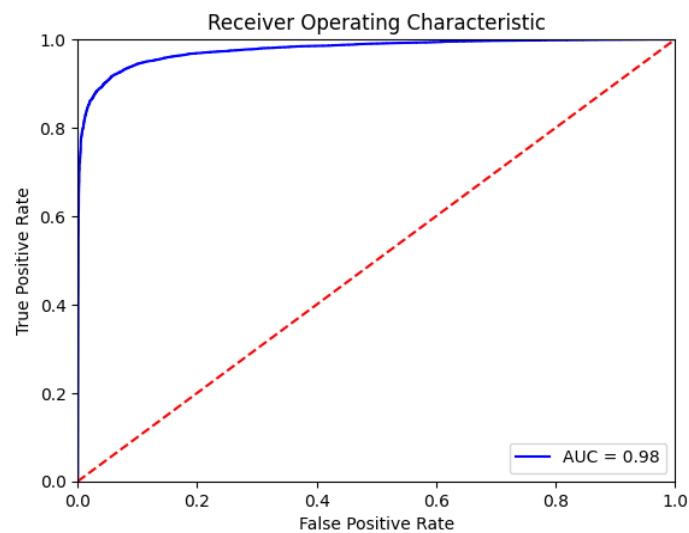

(c) DenseNet121

*Figure 11: ROC Curves for CNN Models on Augmented Dataset*

The ROC curves in Figure 11 demonstrate enhanced performance, with curves closer to the top-left corner compared to Phase 1. The AUC values are listed in Table 4.

Table 4: AUC Values for Models Trained on Augmented Dataset

| Model | AUC |
|---|---|
| **VGG16** | 0.96 |
| **InceptionV3** | 0.97 |
| **DenseNet121** | 0.98 |

Table 4 indicates that the improvement in AUC values further supports the effectiveness of data augmentation in enhancing model performance, with DenseNet121 achieving the highest AUC of 0.98. This supports the conclusion that augmenting the dataset introduces beneficial variability, enabling models to learn more robust and generalised features.

## 4.3 Transfer Learning Results (Phase 3)

The third phase involved fine-tuning the pre-trained models from Phase 2 on the non-augmented dataset using the transfer learning methodology. These pre-trained models were loaded from saved files containing no test data information, ensuring no data leakage during transfer learning. This process allows the models to retain the generalisation capabilities learned from the augmented data while adapting to the specific nuances of the non-augmented dataset. The results for this phase are summarised in Table 5.

Table 5: Performance of Transfer Learning Approach on Non-Augmented Dataset

| **Model** | **Accuracy** | **Precision** | **Recall** | **F1-Score** |
|---|---|---|---|---|
| **VGG16** | 0.92 | 0.96 | 0.89 | 0.92 |
| **InceptionV3** | 0.93 | 0.94 | 0.93 | 0.93 |
| **DenseNet121** | 0.94 | 0.98 | 0.91 | 0.95 |

Table 5 indicates that transfer learning significantly boosted the model performance compared with the non-augmented and augmented dataset experiments. DenseNet121 achieved the highest accuracy (0.94), thereby demonstrating the effectiveness of transfer learning in this domain. Fine-tuning the pre-trained models allowed them to adapt to the raw, non-augmented dataset without losing the generalisation capabilities gained during training on the augmented data. Figure 3 visually compares the model performance and confusion matrices for the transfer learning results.

**Loss Accuracy Curve from Transfer Learning**

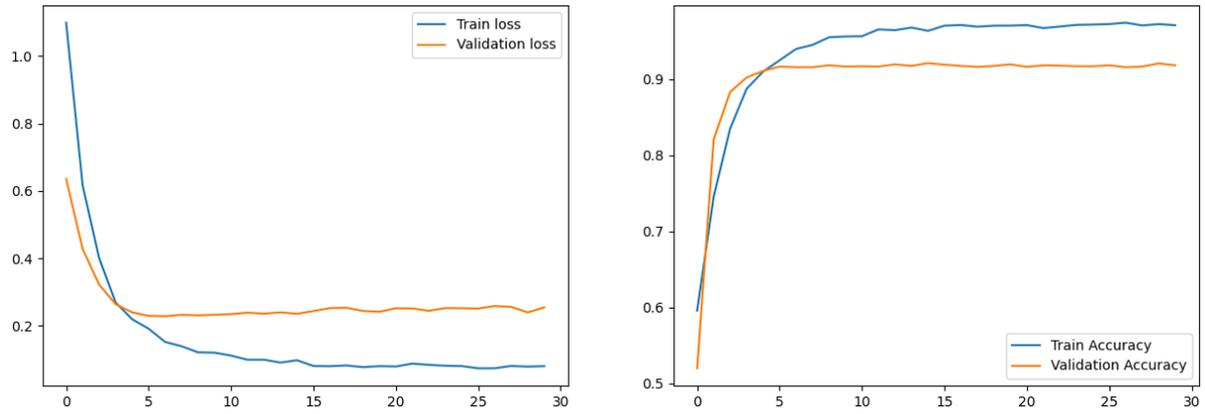

(a) VGG16

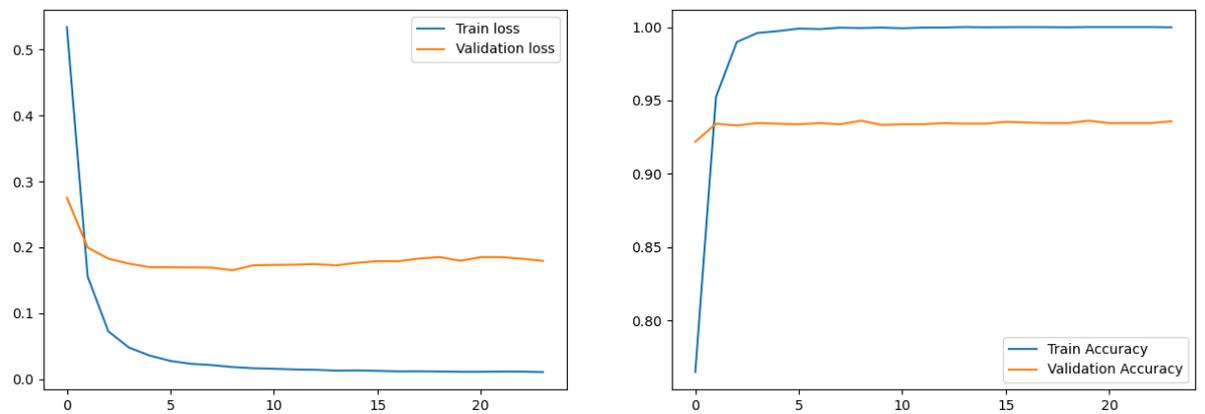

(b) InceptionV3

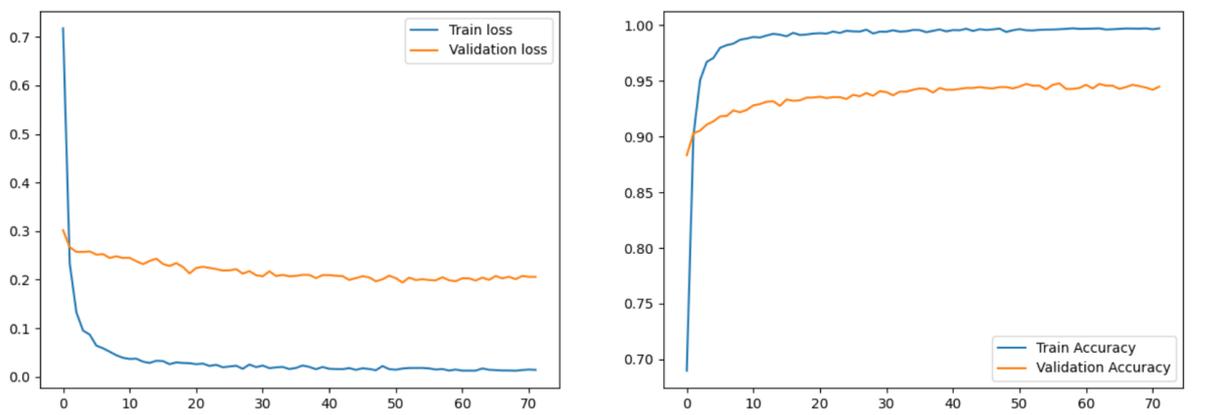

(c) DenseNet121

*Figure 12: Training-Validation Loss and Accuracy Curves after Transfer Learning*

In Figure 12, the models show rapid convergence with lower losses and higher accuracies compared to Phase 1 and Phase 2, indicating that transfer learning enhanced the model's performance on the non-augmented dataset.

**Confusion Matrices from Transfer Learning**

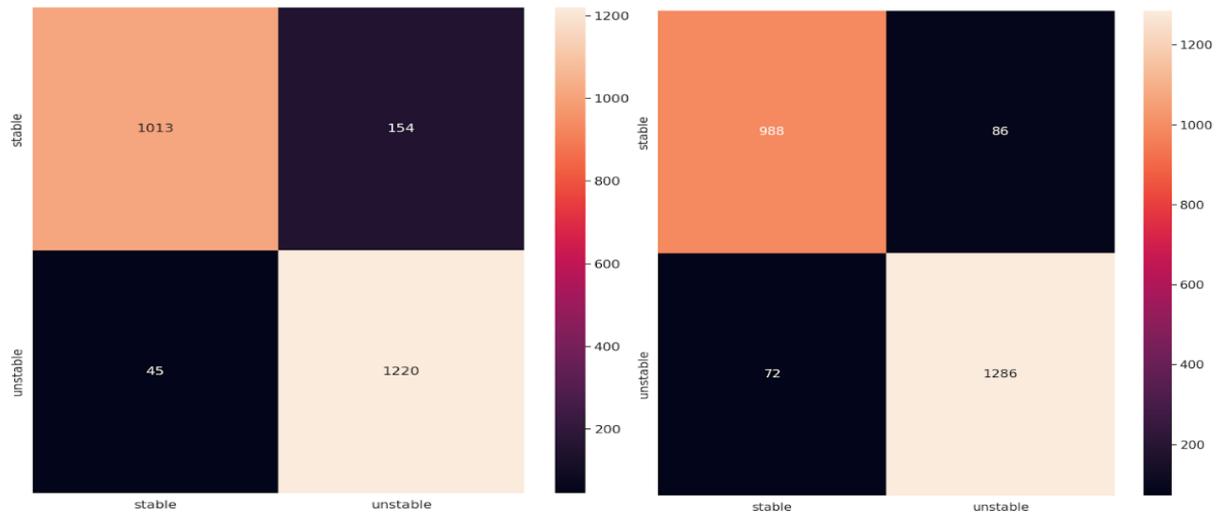

(a) VGG16   (b) InceptionV3

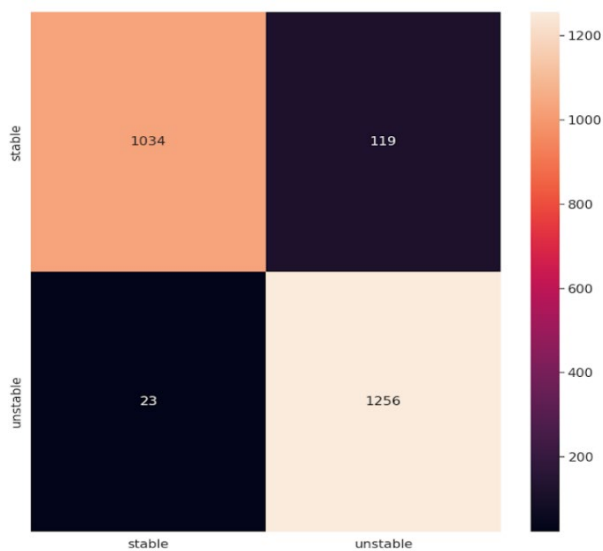

(c) DenseNet121

*Figure 13: Confusion Matrices for CNN Models after Transfer Learning*

The confusion matrices in Figure 13 reveal that DenseNet121 correctly classified almost all samples, significantly reducing the misclassification rates compared with the previous phases.

**ROC Curves from Transfer Learning**

Additionally, the ROC curves and AUC values for the transfer learning approach showed substantial improvements, as shown in Figure 14.

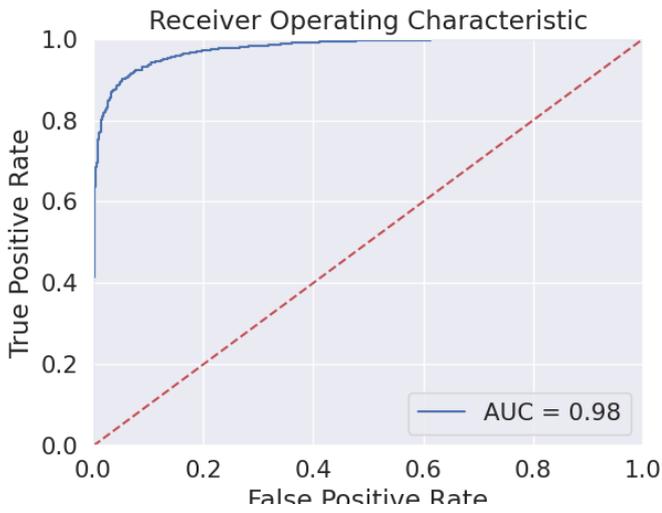

(a) VGG16

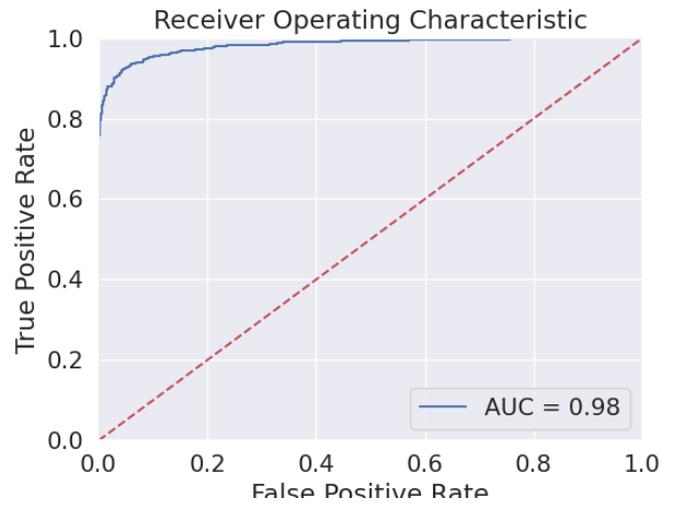

(b) InceptionV3

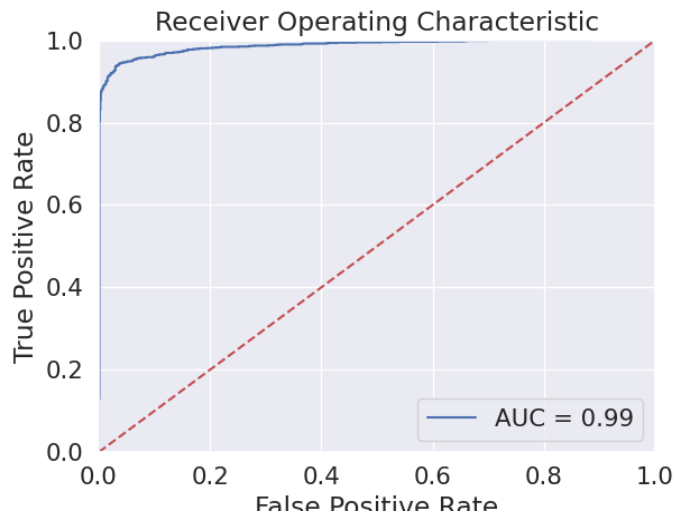

(c) DenseNet121

*Figure 14: ROC Curves and AUC from the Transfer Learning Approach*

The ROC curves in Figure 14 are closest to the top-left corner among all the phases, indicating excellent discrimination ability. The AUC values are listed in Table 6.

Table 6: AUC Values for Transfer Learning Approach

| Model | AUC |
| --- | --- |
| **VGG16** | 0.98 |
| **InceptionV3** | 0.98 |
| **DenseNet121** | 0.99 |

As indicated in Table 6, transfer learning substantially improved the AUC scores across all models, with DenseNet121 achieving an AUC of 0.99. This demonstrates the potential of the proposed transfer learning approach to enhance the classification performance in tasks with limited data, such as mental stability detection using voice spectrograms.

## 4.4 Analysis of Experimental Results

A comparative analysis of the three experimental phases was conducted to better understand the significance of the improvements achieved through data augmentation and transfer learning. Table 7 summarises the results for each model in terms of accuracy, precision, recall, and F1 score across the non-augmented, augmented, and transfer-learning experiments.

Table 7: Comparative Performance of CNN Models Across Different Phases

| **Model** | **Phase** | **Accuracy** | **Precision** | **Recall** | **F1-Score** |
|---|---|---|---|---|---|
| **VGG16** | Non-Augmented | 89% | 96% | 84% | 90% |
| **VGG16** | Augmented | 90% | 93% | 90% | 91% |
| **VGG16** | Transfer Learning | 92% | 96% | 89% | 92% |
| **InceptionV3** | Non-Augmented | 88% | 94% | 84% | 89% |
| **InceptionV3** | Augmented | 91% | 94% | 90% | 92% |
| **InceptionV3** | Transfer Learning | 93% | 94% | 93% | 93% |
| **DenseNet121** | Non-Augmented | 92% | 96% | 89% | 92% |
| **DenseNet121** | Augmented | 93% | 95% | 92% | 94% |
| **Densnet121** | Transfer Learning | 94% | 98% | 91% | 95% |

By analysing Table 7, we assessed the performance improvements of three convolutional neural network models (VGG16, InceptionV3, and DenseNet121) across Non-Augmented, Augmented, and Transfer Learning datasets, focusing on key metrics such as accuracy, precision, recall, F1-score, and AUC. For VGG16, transitioning from the non-augmented to the augmented dataset led to a **1.1% increase** in accuracy (from 89% to 90%) and a significant 7.1% improvement in recall (from 84% to 90%), although there was a **3.1% decrease** in precision (from 96% to 93%). Advancing Transfer Learning further **enhanced accuracy by 2.2%** (reaching 92%) and precision by **3.2%** (up to 96%), whereas recall **slightly declined by 1.1%** (down to 89%). InceptionV3 exhibited a **3.4% gain in accuracy** (88–91%) and a **7.1% rise** in recall (84–90%) when utilising the augmented dataset, with the precision remaining constant at 94%. The shift to Transfer Learning yielded additional improvements for InceptionV3, with **accuracy increasing by 2.2%** (93%) and **recall by 3.3%** (93%). DenseNet121 consistently demonstrated superior performance, showing a **1.1% increase** in accuracy (from 92% to 93%) and a **3.4% boost** in recall (from 89% to 92%) from the Non-Augmented to Augmented datasets, despite a minor **1.0% decrease** in precision (from 96% to 95%). Transitioning to Transfer Learning, DenseNet121's accuracy **rose by another 1.1%** (to 94%), and precision **improved** markedly by 3.2% (to 98%), although recall experienced a slight **reduction of 1.1%** (to 91%). These findings validate the effectiveness of data augmentation and Transfer Learning in improving deep learning models and suggest that DenseNet121 with Transfer Learning is the most efficacious configuration tested. Simultaneously, InceptionV3 with Transfer Learning may be preferred in scenarios where maximising recall is paramount.

The results underscore the critical role of data augmentation and transfer learning in improving the model's robustness and generalisation capabilities. By pre-training on augmented datasets and fine-tuning on non-augmented datasets, the models can learn generalised features that can effectively adapt to the specific characteristics of real-world data. This approach improves classification accuracy and mitigates the risk of overfitting, which is a common challenge in machine learning tasks involving limited data.

## 5. Discussions

The results of this study revealed the effectiveness of the proposed transfer learning methodology in enhancing the classification of mental stability using voice spectrograms. This approach, which leverages pre-training on augmented datasets followed by fine-tuning on non-augmented datasets, demonstrates significant improvements in model performance. This section discusses the implications of the results, compares them with those of relevant studies, and addresses their broader impact on mental health diagnostics.

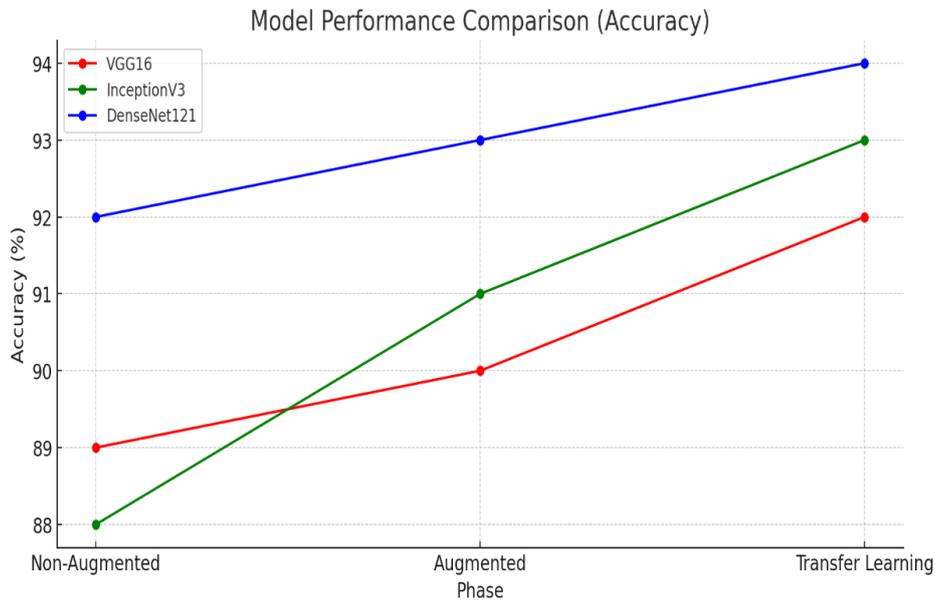

(a) Performance Comparison Graph (Accuracy)

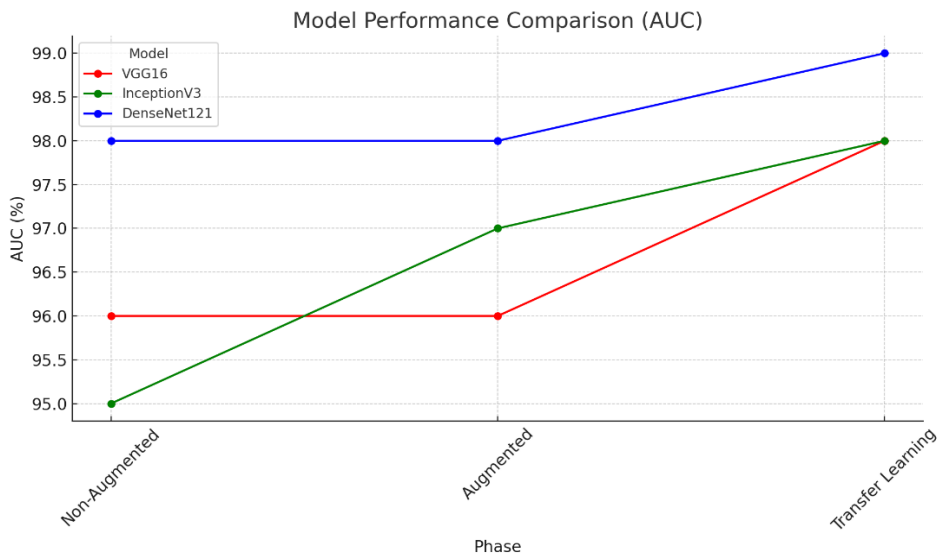

(b) Performance Comparison Graph (AUC)

Figure 15: The performance comparison graphs

The performance comparison graphs (Figure 15 (a)) illustrate the significant improvements in accuracy across the three phases (Non-Augmented, Augmented, and Transfer Learning) and highlight the substantial impact of the proposed methodologies. In the non-augmented phase, DenseNet121 showed the highest accuracy (approximately **92%**), followed by VGG16 (approximately **89%**) and InceptionV3 (**88%**). As the models transitioned to the augmented phase, all three models exhibited improvements, with DenseNet121 continuing to lead, reaching approximately **93%**. InceptionV3 improved to approximately **91%**, whereas VGG16 gradually increased to > **90%**. The Transfer Learning phase further enhanced the performance of all the models, with DenseNet121 achieving the highest accuracy of over **94%**, InceptionV3 closing at **93%**, and VGG16 just above **92%**. This trend indicates that DenseNet121 consistently outperformed the other two models throughout each phase, with InceptionV3 catching up more rapidly, whereas VGG16 showed slower but steady improvements across the phases.

The AUC graph (Figure 15 (b)) shows that during the non-augmented phase, DenseNet121 performed the best, with an AUC of 98%, whereas InceptionV3 had the lowest AUC at 95%. In the Augmented phase, DenseNet121 maintained an AUC of 98%, InceptionV3 showed the largest improvement, rising to 97%, and VGG16 remained constant at 96%. In the Transfer Learning phase, all models improved, with DenseNet121 achieving the highest AUC of 99%, followed by VGG16 and InceptionV3, reaching 98%. The key finding from this graph is that the proposed transfer learning technique provides the most significant performance boost across all models, with DenseNet121 consistently outperforming the others in each phase.

The application of convolutional neural networks (CNNs) to voice spectrograms for mental health classification has proven effective in this study, which aligns with prior research. For instance, Kim et al. (2022) demonstrated the viability of using CNNs and log-mel-spectrograms to detect major depressive disorder (MDD). Their results, similar to ours, demonstrated the ability of CNNs to differentiate between patients and healthy individuals. In this study, among the three CNNs, DenseNet121 achieved the highest classification accuracy of 94% and an AUC score of 99%. These findings further support the evidence that CNN architectures are well suited for tasks requiring the extraction of nuanced patterns from complex datasets, particularly voice spectrograms.

Data augmentation played a crucial role in this study by enhancing model generalisation. Techniques such as SpecAugment, random erasing, and Gaussian noise introduce variability into the training data, allowing CNN models to learn more generalised representations. This is particularly effective when combined with transfer learning. Unlike traditional transfer-learning methods, which typically involve fine-tuning pre-trained models from unrelated tasks, our approach leverages augmented and non-augmented versions of the same dataset to transfer knowledge. The results indicate that this method improves the classification accuracy and ensures that the models maintain their performance when exposed to various real-world data conditions. This integrated approach underscores the potential of combining data augmentation with transfer learning to develop more resilient and accurate models for mental stability classification using voice spectrograms.

A key component of the transfer learning process is the fine-tuning of models on a non-augmented dataset. During fine-tuning, the lower layers of the CNNs that learned general features from the augmented dataset were frozen to retain the learned features. Only the upper layers were fine-tuned, allowing the models to adapt to specific patterns of the non-augmented dataset. This approach is similar to that of Saini et al. (2023), who found that fine-tuning pre-trained CNNs, such as VGG16, on speech emotion detection tasks led to improved classification accuracy by preserving the learned features while adapting to new tasks. In our study, the fine-tuning process resulted in marked improvements across all models, particularly DenseNet121, which maintained the highest classification performance.

The data-split technique for the training and evaluation datasets ensures that the performance metrics of the models are valid and not inflated by data leakage, which is a critical concern in transfer learning. As emphasised by previous studies on machine learning, including Hodgetts et al. (2021), avoiding data leakage is essential to ensure that the model's generalisation performance is accurately measured. This study ensured scientific rigour and transparency by separating the augmented and non-augmented datasets during training and evaluation.

The broader implications of these findings extend beyond mental stability classification. The success of transfer learning in this study suggests that similar approaches can be applied to other healthcare domains where labelled data are scarce. Voice spectrogram analysis, in particular, offers a non-invasive means of diagnosing mental health conditions, making it an attractive tool for clinicians. The classification accuracy and generalisation improvements observed in this study demonstrate that, when combined with data augmentation and transfer learning, CNN models can handle complex audio data in healthcare settings.

The findings of this study align with those of previous studies on mental health diagnostics that used deep learning and voice spectrograms. The combination of data augmentation and transfer learning proved effective in enhancing the model performance, with DenseNet121 emerging as the most robust model. This innovative approach of pre-training augmented data and fine-tuning non-augmented data represents a significant advancement in voice-based mental health classification. Future research could extend this methodology to other healthcare-related tasks, particularly those involving small datasets, further broadening the applicability of deep learning in clinical diagnostics.

## 6. Ethical Considerations

Research using voice signals creates ethical challenges because of the sensitive nature of data. Voice recordings can reveal personal information, including emotional states and identifiable characteristics. Protecting the privacy of participants is crucial. All voice data in this study were anonymised to prevent the identification of individuals. Informed consent will be obtained, and participants will be informed of the specific purposes of the study, methods of data handling, and potential uses of their data for future research. Compliance with institutional review boards (IRBs), which mandate strict controls on collecting, storing, and sharing personal data. In alignment with these principles, any data sharing in this study will follow secure protocols to prevent unauthorised access and misuse.

One of the critical risks in machine learning research, particularly in transfer learning, is data leakage—the inadvertent use of test data during the model training phase–leading to overly optimistic performance estimates. Strict separation of the training, validation, and test datasets was maintained throughout all experimental phases to prevent data leakage. The pre-training phase for the augmented dataset used only the training data, whereas separate validation and test sets were used exclusively for model tuning and evaluation, respectively.

Transparency in the research methodology was maintained through detailed data preparation and model evaluation documentation. The codebase, pre-trained models, and anonymised datasets (where possible) were shared in open-access repositories to facilitate replication and validation by other researchers.

## 7. Conclusion and Future Work

This study introduces a novel transfer learning methodology to enhance the classification of mental stability using voice spectrograms via convolutional neural networks (CNNs). The proposed approach utilises data augmentation techniques, such as SpecAugment and Gaussian noise, to address the challenges posed by limited labelled datasets. By pre-training the CNNs on augmented datasets and fine-tuning them on non-augmented datasets, the models demonstrated significant improvements in classification performance, achieving an accuracy of 94% and an AUC score of 0.99, particularly in the DenseNet121 architecture. This transfer learning approach effectively mitigated overfitting, which is a common issue in models trained on small datasets.

Despite these promising results, this study had several limitations. The implementation of this transfer learning technique primarily focuses on spectrogram images derived from voice signals. Although the proposed methodology yields promising results in this specific context, it remains uncertain whether this approach will perform efficiently on other less popular domain datasets, particularly those with limited data availability, such as the dataset used in this study.

Future work should explore the applicability of this transfer learning methodology across various datasets, particularly those that are less explored or have limited labelled data. Researchers could also investigate how data augmentation and transfer learning principles can be adapted to enhance model performance in these alternative contexts.

This study demonstrates the effectiveness of combining data augmentation and transfer learning to improve the CNN-based classification of mental stability using voice spectrograms. The success of this method has broad implications for AI applications in healthcare, particularly in non-invasive mental health diagnostics. Voice spectrogram analysis, enhanced by convolutional neural networks (CNNs) and transfer learning, offers a scalable tool for the early detection of mental health conditions, complementing traditional assessment methods. As mental health diagnostics increasingly adopt AI-driven approaches, integrating advanced deep learning techniques, such as those explored in this study, will be pivotal in shaping the future of healthcare.